# Fabrication and characterization of high quality factor silicon nitride nanobeam cavities

**Mughees Khan,* Thomas Babinec, Murray W. McCutcheon, Parag Deotare, and Marko Lončar**

*School of Engineering and Applied Sciences, Harvard University, Cambridge, MA 02138, USA*

*Corresponding author: mkhan@seas.harvard.edu*

$Si_3N_4$ is an excellent material for applications of nanophotonics at visible wavelengths due to its wide bandgap and moderately large refractive index ($n \approx 2.0$). We present the fabrication and characterization of $Si_3N_4$ photonic crystal nanobeam cavities for coupling to diamond nanocrystals and Nitrogen-Vacancy centers in a cavity QED system. Confocal micro-photoluminescence analysis of the nanobeam cavities demonstrates quality factors up to $Q \sim 55{,}000$, which is limited by the resolution of our spectrometer. We also demonstrate coarse tuning of cavity resonances across the 600-700nm range by lithographically scaling the size of fabricated devices. This is an order of magnitude improvement over previous $SiN_x$ cavities at this important wavelength range. 

Visible optical microcavities have a wide variety of applications, ranging from classical and quantum information processing to compact biological and chemical sensing. Many recent proposals for the development of a solid-state quantum information processing system have focused on visible nanophotonic devices with integrated diamond color centers due to their photostability and room temperature operation. Additional emphasis is placed on the Nitrogen-Vacancy (NV) center as an active element since it possesses both spin[1-4] and photon[5-7] quantum bits that can be optically initialized and read-out. Recent developments, such as diamond nanowire antennas fabricated from bulk diamond samples[7-8], plasmon-enhanced antennas[9], and optical microcavities in other semiconductor material systems[10-14] coupled to proximal diamond nanocrystals, have shown that it is possible to engineer the optical properties (e.g. collection efficiency, single photon generation rate) of a single NV center. An alternative, noteworthy system that has been shown to theoretically approach the strong-coupling regime of cavity quantum electrodynamics (cQED) is based on coupling the zero-phonon line emission (637 nm) to a high quality factor ($Q \sim 10^5$) silicon nitride ($SiN_x$) nanobeam cavity[15]. Towards this end, we report in this paper on the fabrication and characterization of a high Q-factor nanobeam photonic crystal (PhC) cavity in an air-bridge $Si_3N_4$ structure. Devices are demonstrated with quality factor $Q \sim 55{,}000$, which is an order of magnitude higher than previously reported at visible wavelengths[16-20] and approaches the regime necessary for such cQED studies.

In a nanobeam PhC cavity, optical confinement is provided by photonic crystal mirrors along the waveguide dimension and by total internal reflection in the other two transverse dimensions. The cavity design for the devices studied in this work was optimized for a 200 nm thick stoichiometric $Si_3N_4$ device layer (n = 2.0). The nanobeam was 300 nm wide and was patterned with a one dimensional photonic crystal lattice of circular holes with periodicity a = 250 nm and radius r = 70 nm, and the spacing between photonic mirrors was chosen to generate a cavity resonance at 637 nm. In order to minimize light scattering outside the cavity, the PhC hole mirror was adiabatically tapered[15,21-23] by linearly reducing the PhC hole spacing from a = 250 nm and hole size r = 70 nm in the mirror to $a_0$ = 205 nm and $r_1$ = 55 nm at the cavity center. Figure 1a shows the cavity mode profile for this 4-hole taper cavity, whose theoretical quality factor $Q$ = 230,000 and mode volume $V_m = 0.55\ (\lambda/n)^3$. The cavity Q factor is highly sensitive to the cavity length, defined as the center-to-center distance of the two central holes, and varies by two orders of magnitude over a 15 nm range.

A high-stress, low-pressure chemical vapor deposition (LPCVD) $Si_3N_4$ film (200nm thickness) on a <100> Si substrate was used in the device layer. A Woollam spectroscopic scanning ellipsometer confirmed the refractive index n ≈ 2.0 of the stoichiometric nitride film prior to processing. The wafers were then solvent cleaned, $N_2$ blow dried and dehydration-baked, and further cleaned in a barrel etcher (Technics Micro-stripper) for 5 min at 100 W power, 200 mT pressure, and 20 sccm oxygen flow rate. Approximately 250nm thick ZEP 520A or PMMA 950C3 was used as electron beam resist, which was spun at 4000 rpm for 40 s and then soft baked at 180º C for 2 min and 3 min, respectively. PMMA showed better adhesion to $Si_3N_4$ compared to ZEP, though the oxygen plasma cleaning step was observed to improved ZEP adhesion. The ELS-7000 (Elionix Inc., Japan) 100 KV electron beam lithography tool was used to pattern the designed 4-hole tapered PhC nanobeam structure in the resist. The ZEP coated samples were developed in O-xylene for 120s, and the PMMA coated samples were developed in MIBK:IPA (1:3) for 90s. After development, the patterned PhC structure was transferred to the $Si_3N_4$ film in a reactive ion etcher (RIE) using a $C_4F_8/SF_6/H_2$ recipe in an STS inductively coupled plasma (ICP) RIE at 120 nm/min $Si_3N_4$ etch rate with smooth, near-vertical sidewalls. We observed a selectivity of 1:1.5 to ZEP and about 1:1.2 to PMMA. PMMA stripping was done in acetone. After stripping the resist, KOH:$H_2O$ (1:4) solution[24-25] was used to etch away the exposed Si at 65ºC. KOH:$H_2O$ also stripped any remaining ZEP. KOH selectivity to different Si planes was considered as part of the fabrication process. This allowed the successful release of the air-bridge PhC cavity, as shown in Figure 1b. The presence of a highly stressed $Si_3N_4$ film allowed releasing of the air-bridged cavity using wet isotropic etching of Si without the use of any critical point drying to release the suspended structures. However, stresses in the LPCVD $Si_3N_4$ film may have caused preferential KOH etching of $Si_3N_4$ around the air holes, making them slightly elliptical[26].

A home-built micro-photoluminescence (μPL) system was used to characterize the low-level, intrinsic fluorescence of the devices. The nanobeam photonic crystals were pumped with ~500 μW of a 532 nm CW laser (Coherent Compass, 315M) using a 100X, 0.95 NA objective. Increasing the pump beam power beyond 500 μW led to higher background fluorescence levels and also damaged the cavities in some cases, potentially due to lack of heat dissipation in the 1D geometry of the nanobeam. A 3-axis piezoelectric stage (Piezosystem Jena, Tritor 100) scanned the sample while the pump beam was fixed. Fluorescence was collected back through the objective and focused on a 1 x 2 single mode fiber beam splitter, which acted as a confocal pinhole. One arm of the beam splitter was connected to an avalanche photodiode (Perkin Elmer) to generate an image of the sample and optically address individual nanobeam devices (Fig. 2b). The second arm of the beam splitter was connected to a spectrometer (Jobin Yvon, iHR 550) in order to identify resonant features in the fluorescence.

The photoluminescence spectrum of a typical device is shown in Figure 3a. A low-resolution (~150lines/mm) grating was used for initial characterization of the device and resulted in artificial broadening of the feature. Still, the high sensitivity of this measurement to the low light levels emitted from the device (~40,000 total photon counts per second) allowed us to confirm the dipole character of the cavity resonance via measurements of its spectrum as a linear polarizer was rotated in the collection path of the setup. High transmission of the cavity signal was observed when the analyzer was parallel to the cavity dipole (Fig. 3a, black) and extinction was observed for the orthogonal direction (Fig. 3a, purple). Moreover, the cavity signal was observed to vanish when taking photoluminescence spectra several spot sizes ~1-2 μm away from the cavity center (Fig. 3a, green). This demonstrates the sensitivity of the cavity resonance to the excitation location, which is expected due to the small mode volume design of the cavity. Once the resonant feature was identified, we then switched to a high-resolution grating (~1800 lines/mm) in order to identify the cavity Quality factor ($Q$). Devices were routinely observed with $Q > 10^4$, though with some variation due to fabrication tolerances. The best device that we observed (Fig. 3b) possessed a quality factor $Q$ ~5.5 x $10^4$, which represents a record for $SiN_x$ photonic crystal cavities operating at visible wavelengths.  Some devices demonstrated even narrower resonances (data not shown), but with few (< 3) points so that the cavity $Q$ factor is difficult to infer from a Lorentzian fit. Finally, we scaled the nanobeam device parameters at a fixed value of r/a in order to shift the resonant wavelength[27-28]. Figure 4 shows good agreement between simulated and measured device wavelengths for cavities with (-2, +2, +5, +10)% scaling. In the future, this could technique could provide a coarse tuning mechanism for coupling to emitters with narrow emission lines.

In this work we have presented the design, fabrication, and characterization of silicon nitride nanobeam cavities at visible wavelengths.  By utilizing a 4-hole taper design, devices with $Q$ ~55,000 and approaching $10^5$ were observed using μPL measurements. An important observation was the difficulty in characterizing cavities with an ultra-high $Q$ factor, which is consistent with our previous results obtained using Si cavities and a resonant scattering setup[29]. Additional characterization of the devices, for example based on a fiber-taper probe or a resonant scattering with a tunable laser system, such as those used elsewhere[30], could allow for the observation of even higher-$Q$ modes. Moreover, the introduction of light emitters such as diamond color centers will allow us to investigate cQED phenomena.

The authors thank the financial support from NSF under NIRT grant ECCS-0708905 and Harvard's National Science and Engineering Center (http://www.nsec.harvard.edu). M. W. McCutcheon kindly thanks NSERC for its generous support. T. Babinec was funded by the NDSEG graduate student fellowship. Most of the nanofabrication work was performed at Center for Nanoscale (CNS) at Harvard University. The authors would also like to thank the CNS staff members, especially Dr. Ling Xie and Steve Paolini for all their help and support.

## Figure Captions

Fig.1. a) Mode profile ($E_y$) of a nanobeam photonic crystal cavity with $Q$ = 230,000 and $V_m \sim 0.55\ (\lambda/n)^3$ based on a 4-hole taper with radius (r) and pitch (a) linearly increasing from 55 to 70nm and 205 to 250nm, respectively, in the mirror sections (starting from the center). b) Fabricated $Si_3N_4$ cavity with arrows that denote the polarization with respect to the cavity.

Fig.2. a) Cartoon of the confocal microscope used in this experiment. b) 2-D confocal microscope image showing an array of cavities having different scaling percentages separated by spacers. Inset shows a zoomed-in image of one cavity.

Fig.3. a) Cavity resonance as a function of polarization measured with a coarse 150 lines/mm grating. b) Typical cavity resonance measured with a high-resolution 1800 lines/mm grating. Data (black circles) and Lorentzian fit (red line) gives $Q \sim$ 55,000.

Fig.4. Comparison between experimental and theoretical data for nanobeam cavities that are scaled versions of the optimal device design.

**Figure 1**

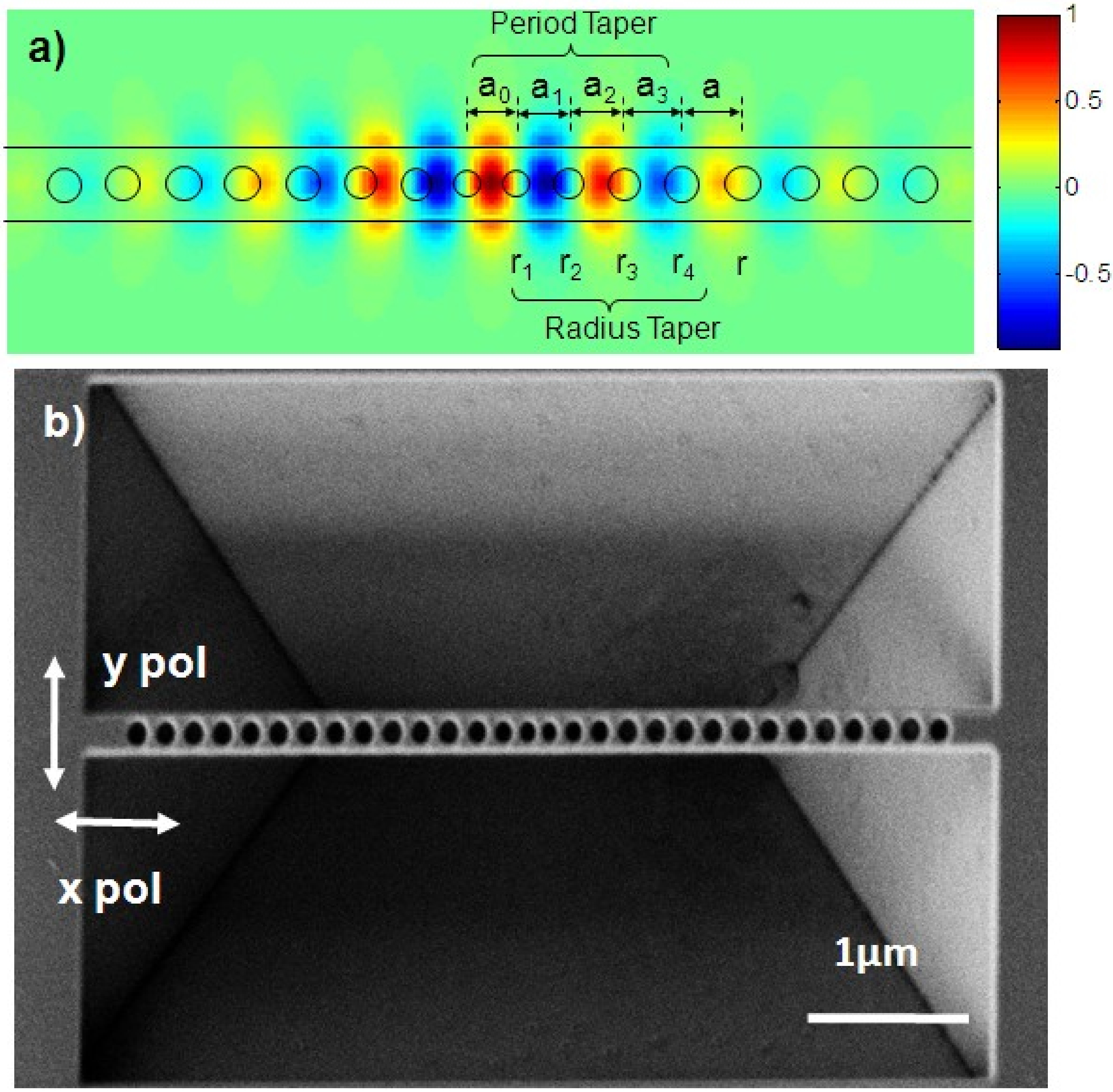
a)
Period Taper
a0 a1 a2 a3 a
r1 r2 r3 r4 r
Radius Taper
1
0.5
0
-0.5
b)
y pol
x pol
1μm

**Figure 2**

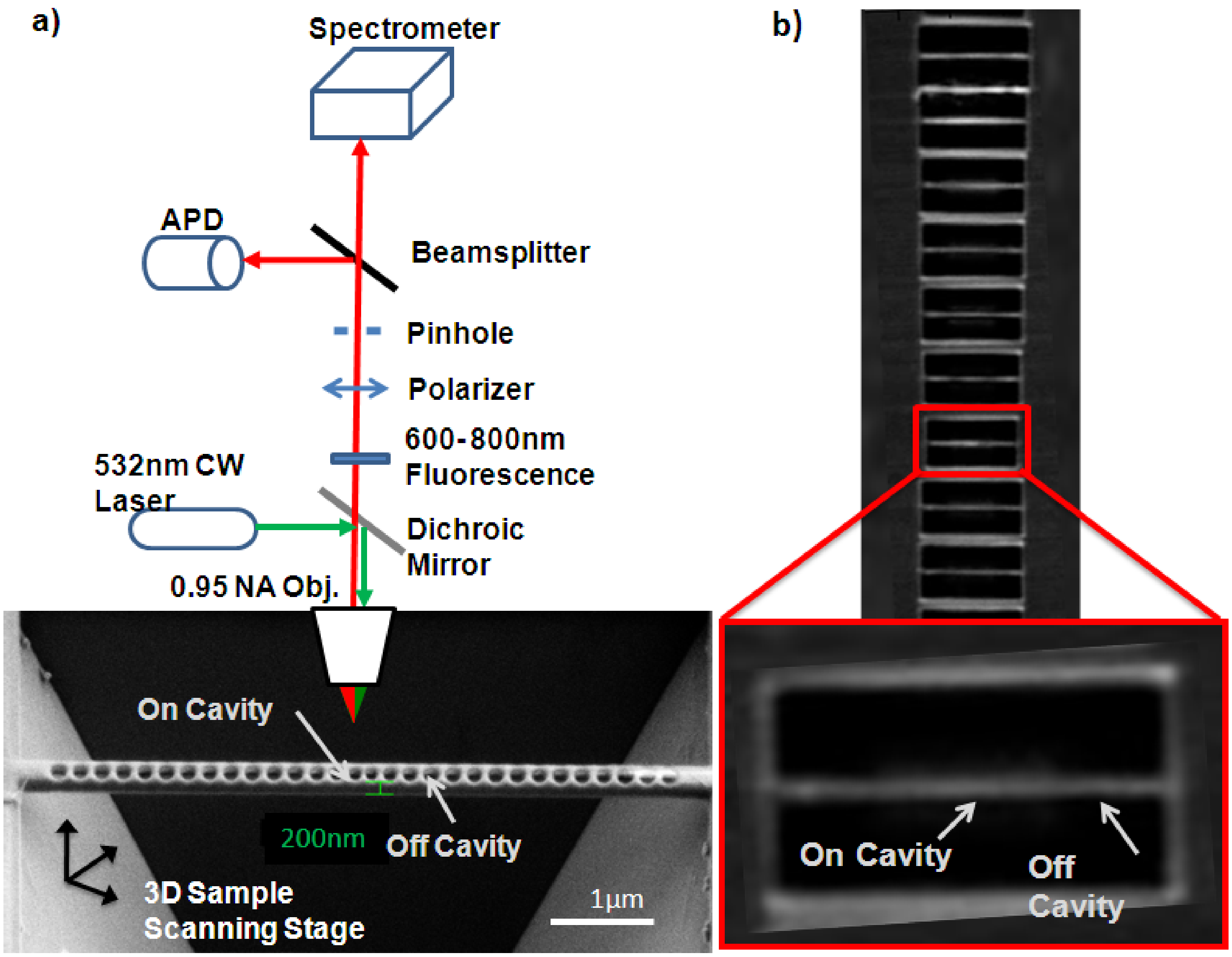
a)
Spectrometer
APD
Beamsplitter
Pinhole
Polarizer
600-800nm
Fluorescence
532nm CW
Laser
Dichroic
Mirror
0.95 NA Obj.
On Cavity
200nm
Off Cavity
3D Sample
Scanning Stage
1µm
b)
On Cavity
Off
Cavity

**Figure 3**

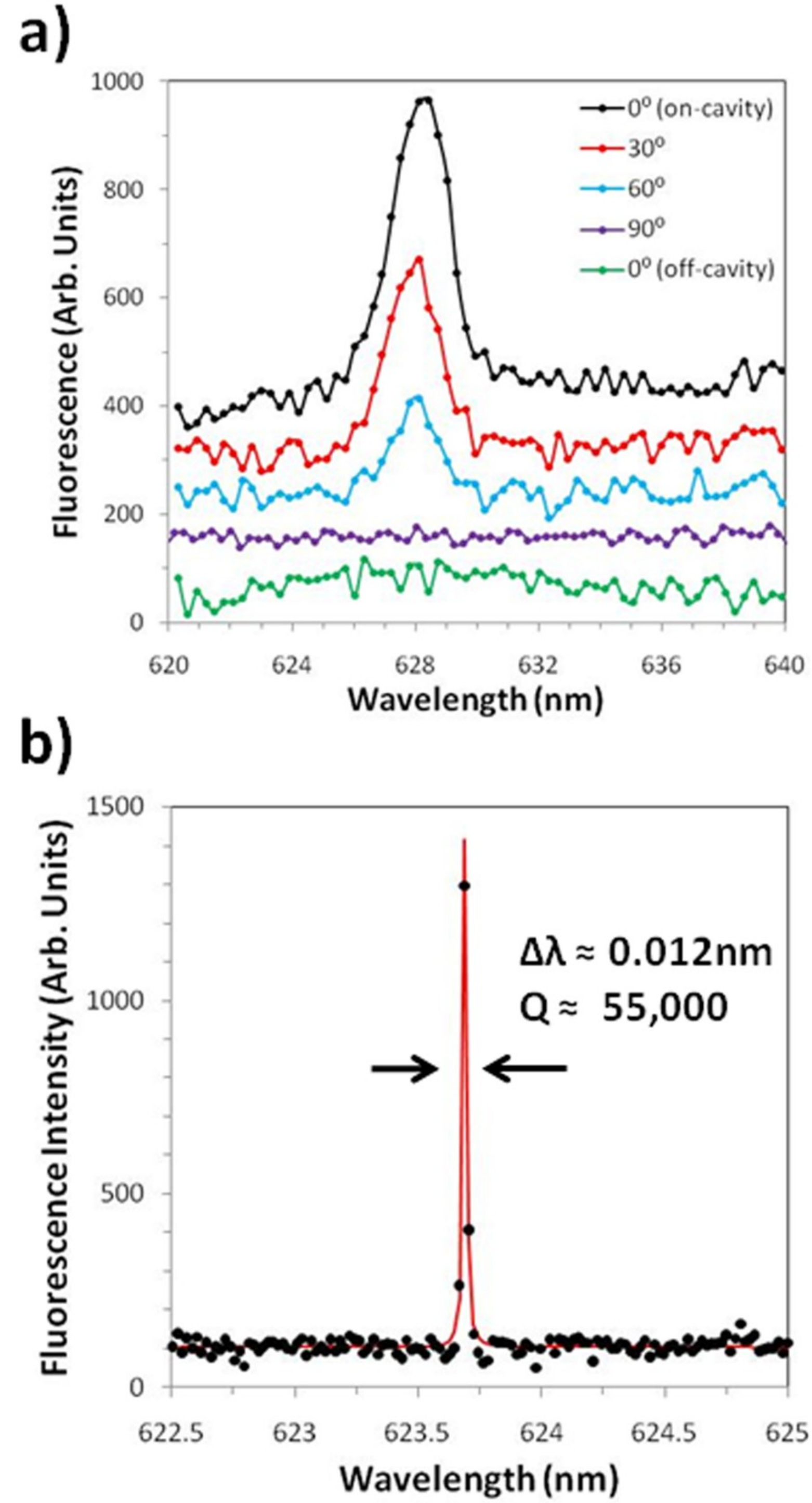
a)
1000
800
600
400
200
0
Fluorescence (Arb. Units)
0° (on-cavity)
30°
60°
90°
0° (off-cavity)
620
624
628
632
636
640
Wavelength (nm)
b)
1500
1000
500
0
Fluorescence Intensity (Arb. Units)
Δλ ≈ 0.012nm
Q ≈ 55,000
622.5
623
623.5
624
624.5
625
Wavelength (nm)

**Figure 4**

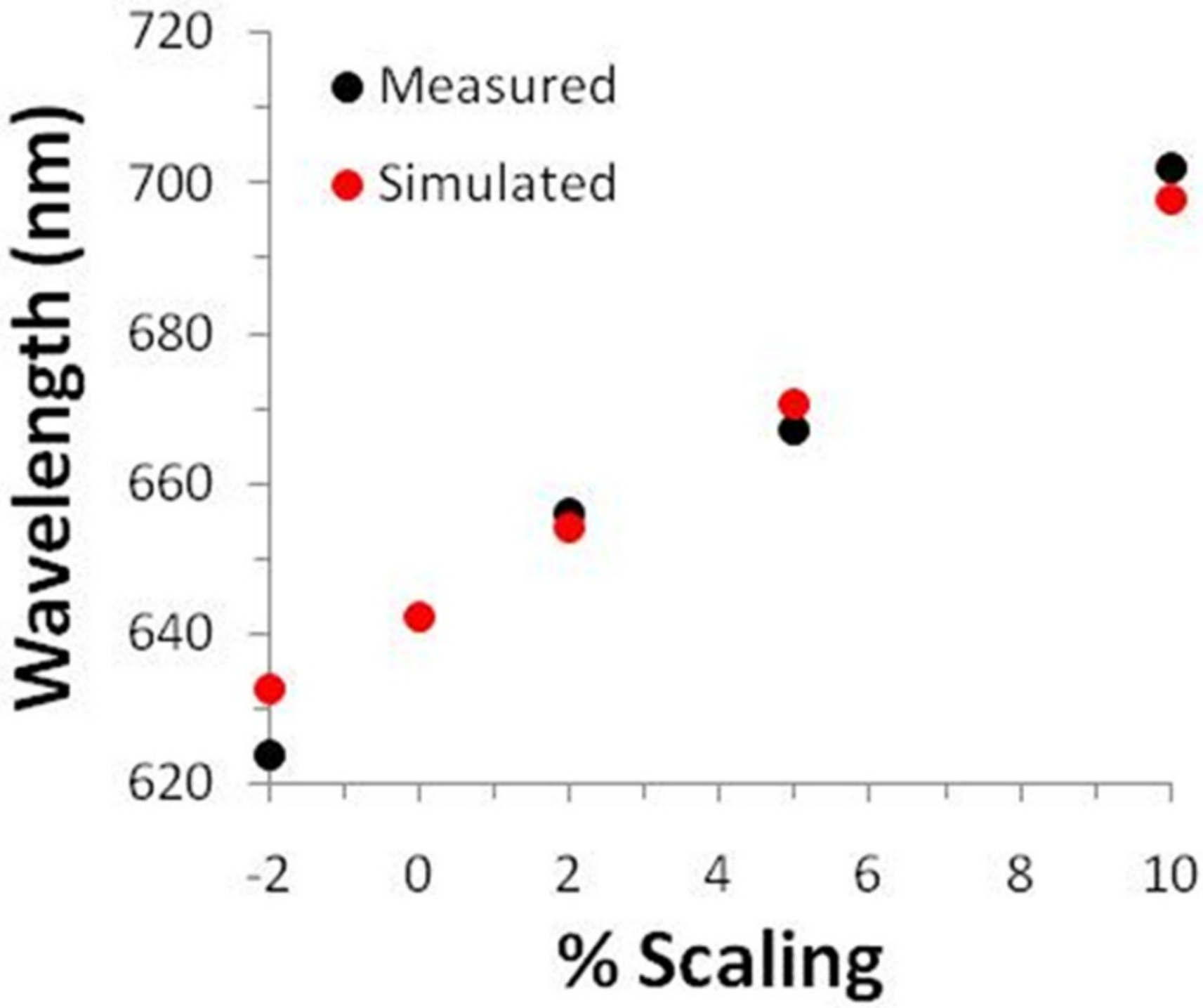

Measured
Simulated
Wavelength (nm)
720
700
680
660
640
620
-2
0
2
4
6
8
10
% Scaling